\documentclass[acus]{JAC2000_mc}
\usepackage{graphicx}
\begin{document}

\thispagestyle{empty}

\onecolumn

\begin{flushright}
{\large
SLAC--PUB--8872\\
June 2001\\}
\end{flushright}

\vspace{.8cm}

\begin{center}

{\LARGE\bf
NLC Beam Properties and Extraction Line Performance\\
with Beam Offset at IP~\footnote
{\normalsize
{Work supported by Department of Energy contract  DE--AC03--76SF00515.}}}

\vspace{1cm}

\large{
Y.~Nosochkov, T.O.~Raubenheimer and K.A.~Thompson\\
Stanford Linear Accelerator Center, Stanford University,
Stanford, CA 94309}

\end{center}

\vfill

\begin{center}
{\LARGE\bf
Abstract }
\end{center}

\begin{quote}
\large{
Properties of the disrupted NLC beam at the Interaction Point (IP) and
particle loss in the extraction line are analyzed as a function of
beam-to-beam position and angular offset at IP.  The simulations show that
disruption and beam loss maximize when the vertical beam separation at IP
is about 20 times the rms vertical beam size.  The horizontal offset does
not increase the disruption and the beam loss.  The angular offsets cause
particle loss in the extraction line mainly because of the beam orbit
oscillations.
}
\end{quote}

\vfill

\begin{center}
\large{
{\it Presented at the 2001 Particle Accelerator Conference
(PAC 2001)\\
Chicago, Illinois, June 18--22, 2001}
} \\
\end{center}

\newpage

\pagenumbering{arabic}
\pagestyle{plain}

\twocolumn

\title
{NLC BEAM PROPERTIES AND EXTRACTION LINE PERFORMANCE\\
WITH BEAM OFFSET AT IP~\thanks
{Work supported by Department of Energy contract
DE--AC03--76SF00515.}\vspace{-6mm} }

\author{Y.~Nosochkov, T.O.~Raubenheimer and K.A.~Thompson\\
SLAC, Stanford University, Stanford, CA 94309, USA}

\maketitle

\begin{abstract}

Properties of the disrupted NLC beam at the Interaction Point (IP) and
particle loss in the extraction line are analyzed as a function of
beam-to-beam position and angular offset at IP.  The simulations show that
disruption and beam loss maximize when the vertical beam separation at IP
is about 20 times the rms vertical beam size.  The horizontal offset does
not increase the disruption and the beam loss.  The angular offsets cause
particle loss in the extraction line mainly because of the beam orbit
oscillations.

\end{abstract}

\vspace{-1mm}
\section{Introduction}

In the NLC~\cite{param}, the strong beam-beam interaction
significantly distorts beam distribution at IP.  This effect, called
disruption~\cite{disrupt}, increases the beam emittance and angular
divergence after collision and generates a huge energy spread in the
outgoing beam.  The NLC extraction line was designed to minimize
particle loss caused by these effects~\cite{dump1,dump2}.  

So far, the NLC beam disruption and extraction line performance were
studied for the ideal beam conditions at the IP.  However, various incoming
beam errors may affect the beam-beam interaction and the resultant
disrupted distribution.  In this paper, we discuss the effects of
beam-to-beam transverse position and angular offsets at IP on the beam
distribution and particle loss in the extraction line.

Beam-beam effects such as energy loss due to beamstrahlung occur as
particles in each bunch go through the strong coherent field of the other
bunch.  When the beams are vertically flat, as they are in NLC, the
beamstrahlung can be enhanced significantly for some vertical position and
angle offsets, because the field seen by the bulk of the particles in one
beam increases as the offset from the other beam increases.  Of course, the
field seen by each beam eventually falls off when position offsets become
large enough.  The interaction is further complicated by the fact that the
two beams distort each other's shape during the collision.  Thus we resort
to numerical simulations, where care must be taken in choosing the grids to
perform beam-beam calculations as accurately as possible.

The beam parameters for the NLC design can be found in Ref.~\cite{param}.
In this study, we used the Stage 1 parameters where the center-of-mass
energy is 500 GeV; the IP parameters listed in Tables~1 and~2, where
$\epsilon$ is the beam emittance, $X_{rms}/Y_{rms}$ the beam size, and
$X^{\prime}_{rms}/Y^{\prime}_{rms}$ angular divergence in the horizontal
and vertical plane, respectively.  The listed parameters are for ideal IP
conditions and zero initial energy spread.  The realistic incoming energy
spread is about $\pm0.4\%$, but it is negligible compared to the disrupted
energy spread and does not affect the results of this study.

\begin{table}[tb]
\vspace{-3mm}
\begin{center}
\caption{NLC parameters in option H.}
\medskip
\begin{tabular}{|l|l|}
\hline
Energy {\it cms} [GeV] & 500 \\
Luminosity [$10^{33}$] & 22 \\
Repetition rate [Hz] & 120 \\
Bunch charge [$10^{10}$] & 0.75 \\
Bunches/RF-pulse & 190 \\
Bunch separation [ns] & 1.4 \\
Eff. gradient [MV/m] & 48 \\
Inject. $\gamma\epsilon_x/\gamma\epsilon_y$ [$10^{-8}$ m-rad] & 300 / 2 \\
Bunch length $\sigma_z$ [$\mu$m] & 110 \\
$\Upsilon_{ave}$ & 0.11 \\
Pinch enhancement & 1.43 \\
Beamstrahlung $\delta_B$ [\%] & 4.6 \\
Photons per $e^+/e^-$ & 1.17 \\
Two linac length [km] & 6.3 \\
\hline
\end{tabular}

\caption{IP parameters before and after collision.}
\medskip
\begin{tabular}{|l|ll|}
\hline
 & before & after \\
\hline
$\gamma\epsilon_x/\gamma\epsilon_y$ [$10^{-8}$ m-rad] 
& 360 / 3.5 & 1175 / 7.2 \\
$\beta_x/\beta_y$ [mm] 
& 8 / 0.1 & 2.44 / 0.14  \\
$\alpha_x/\alpha_y$ 
& 0 / 0 & 1.852 / 0.675 \\
$X_{rms}/Y_{rms}$ [nm] 
& 245 / 2.7 & 245 / 4.6 \\
$X^{\prime}_{rms}/Y^{\prime}_{rms}$ [$\mu$rad] 
& 31 / 27 & 211 / 39 \\
\hline
\end{tabular}
\end{center}
\vspace{-4mm}
\end{table}

\vspace{-1mm}
\section{SIMULATIONS}

In the study, the undisrupted beam parameters in Tables~1,2 were used as
the input data for GUINEA--PIG code~\cite{gpig} to generate the incoming
gaussian beams, simulate beam-beam interaction and obtain disrupted
distribution at the IP.  Some of the disrupted beam parameters for
$5\!\times\!10^4$ particles are shown in Table~2 (after collision), where
the disrupted values of $\beta$, $\alpha$ and $\epsilon$ were derived from
the beam distribution.  The disruption significantly increases beam
emittance and angular divergence and generates huge energy spread with 
low energy tail up to $\frac{\Delta E}{E}\!\sim\!-70$\%.

The disrupted beam was then tracked from IP to the dump to compute particle
loss in the extraction line.  Only the primary beam particles were used in
this simulation.  The tracking was performed using the NLC version of DIMAD
code~\cite{dimad} which correctly accounts for very large energy errors
present in the NLC disrupted beam.

The extraction line optics used in this study is described in
Ref.~\cite{dump2}.  It consists of two multi-quadrupole systems separated
by a four bend chicane with 2~cm vertical dispersion.  The effects of 6~T
detector solenoid were included, but no magnet errors were used.

\vspace{-1mm}
\section{POSITION OFFSET AT IP}

To generate disrupted beam distributions for various values of beam-to-beam
offset at IP, the colliding beams were symmetrically and oppositely
displaced at IP by half of the total offset $\pm\frac{1}{2}\Delta
x$ or $\pm\frac{1}{2}\Delta y$, in the GUINEA--PIG simulations.  The
resultant distributions with $5\!\times\!10^4$ particles were tracked in the
extraction line to compute a beam loss.

For $\Delta x\!=\!\Delta y\!=\!0$, the disrupted parameters are already
given in Table~2.  As the beam offset increases, the disrupted distribution
should eventually converge to the incoming beam distribution since less
interaction takes place, however,
at very large offsets the beam loss in the extraction line may be
caused by the large incoming orbit.

\begin{figure}[b]
\centering
\includegraphics*[width=50mm]{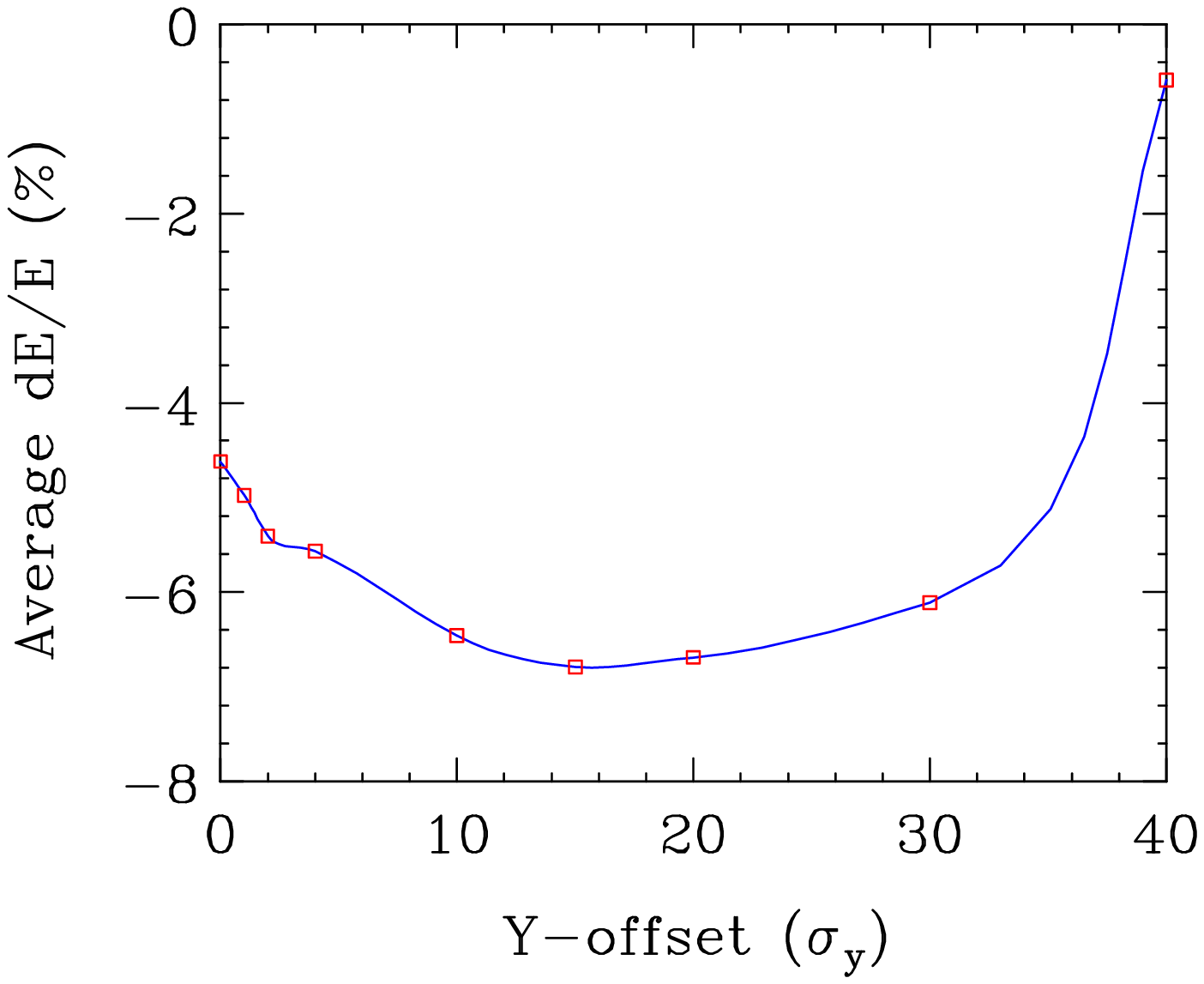}
\vspace{-3mm}
\caption{Average disrupted $\frac{\Delta E}{E}$ at IP vs. $\Delta y$.}
\label{deave-offy-H490}
\vspace{-1mm}
\end{figure}

\begin{figure}[t]
\centering
\includegraphics*[width=50mm]{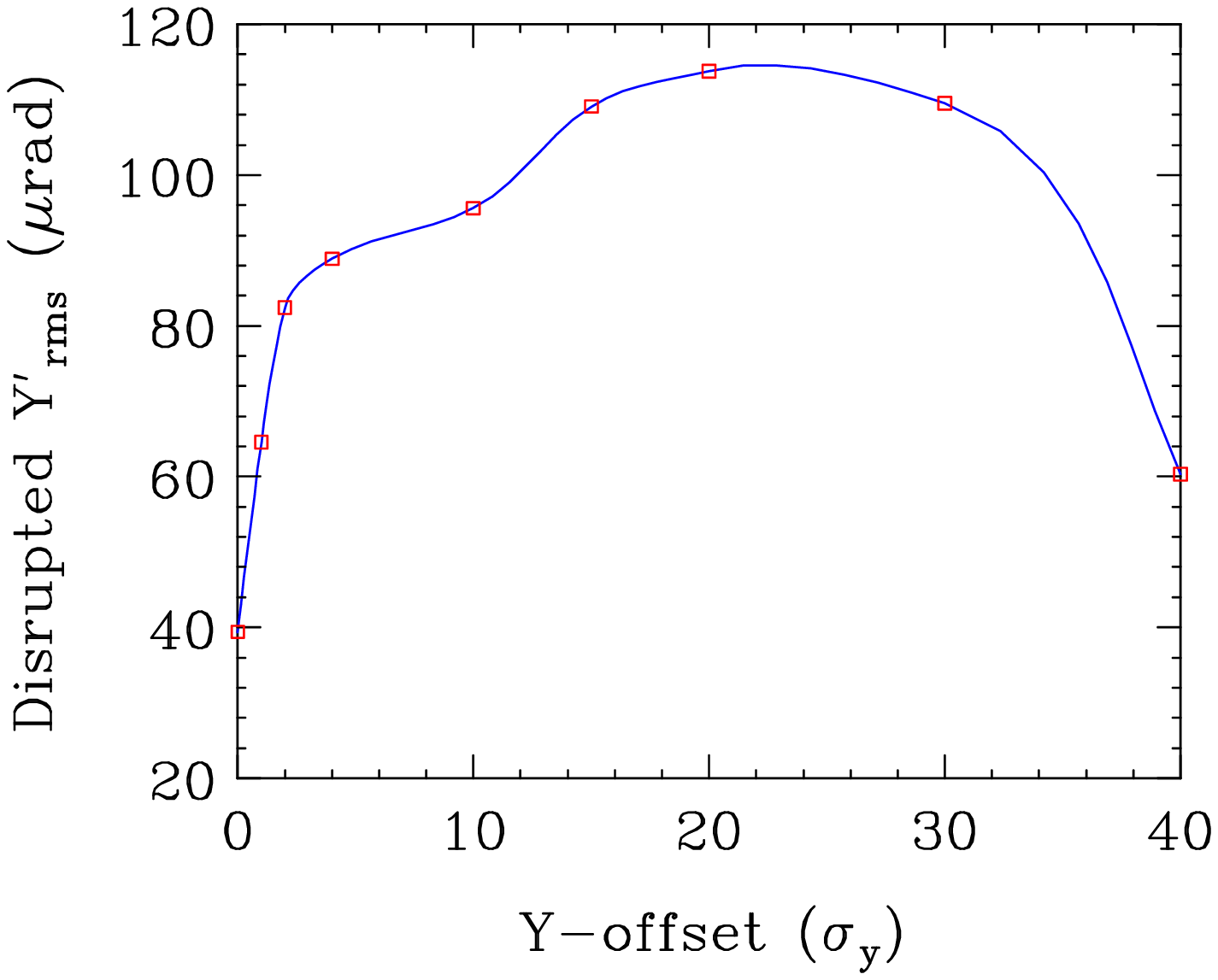}
\vspace{-3mm}
\caption{Disrupted $rms$ vertical divergence at IP vs. $\Delta y$.}
\label{sigpy-offy-H490}
\vspace{-0mm}
\end{figure}

\begin{figure}[h!]
\centering
\includegraphics*[width=70mm]{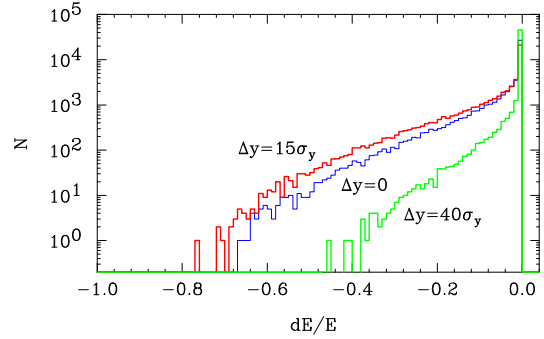}
\vspace{-3mm}
\caption{Energy spread at IP for
$\Delta y\!=\!0$, $15\sigma_y$ and $40\sigma_y$.}
\label{ehist-offy-H490}
\vspace{-0mm}
\end{figure}

A summary of disrupted beam parameters for various beam offsets at IP are
given in Table~3 where $\sigma_{x,y}$ are the undisrupted $rms$ beam size at
IP.  One can see that $x$-offset
gradually reduces the beam disruption, but the $y$-offset initially
increases the energy spread, the vertical beam size and the divergence.  The
disruption maximizes at $\Delta y\!\approx$15 to 20$\sigma_y$, but the
maximums are rather broad as shown in Fig.~\ref{deave-offy-H490}
and~\ref{sigpy-offy-H490}.  Analysis of distributions in Table~3 shows that
vertical emittance is blown up more than ten times at $\Delta
y\!=\!20\sigma_y$.  A comparison of the energy spread at $\Delta y\!=\!0$,
$15\sigma_y$ and $40\sigma_y$ is shown in Fig.~\ref{ehist-offy-H490}.

\begin{table*}[tb]
\vspace{-3mm}
\begin{center}
\caption{Disrupted IP parameters vs. beam offset at IP.}
\vspace{1mm}
\medskip
\begin{tabular}{|rl|cccccc|}
\hline
\multicolumn{2}{|c|}{Offset} & $X_{rms}$ & $X^{\prime}_{rms}$ 
& $Y_{rms}$ & $Y^{\prime}_{rms}$ 
& $(\frac{\Delta E}{E})_{ave}$ & $(\frac{\Delta E}{E})_{rms}$ \\
& & [nm] & [$\mu$rad] & [nm] & [$\mu$rad] 
& [\%] & [\%] \\
\hline
\hline
& 0          & 244.7& 210.8&  4.63&  39.4& -4.62&  8.13 \\
\hline
\hline
$\Delta x=$
& 1$\sigma_x$& 244.9& 157.7&  4.35&  36.9& -4.25&  7.71 \\
& 2$\sigma_x$& 246.5&  90.4&  3.60&  32.0& -3.13&  6.39 \\
& 4$\sigma_x$& 243.0&  62.7&  2.75&  27.4& -0.78&  2.66 \\
\hline
\hline
$\Delta y=$ 
& 1$\sigma_y$& 245.9& 206.8&  8.22&  64.6& -4.98&  8.43 \\
& 4$\sigma_y$& 245.7& 192.4& 11.91&  88.9& -5.57&  9.06 \\
&10$\sigma_y$& 246.5& 202.3& 14.91&  95.6& -6.46&  9.89 \\
&15$\sigma_y$& 244.6& 196.0& 16.23& 109.1& -6.79& 10.13 \\
&20$\sigma_y$& 244.7& 187.1& 17.27& 113.8& -6.69& 10.05 \\
&30$\sigma_y$& 245.1& 167.8& 17.27& 109.5& -6.11&  9.50 \\
&40$\sigma_y$& 244.9&  48.4&  4.94&  60.3& -0.59&  2.49 \\
\hline
\end{tabular}
\end{center}
\vspace{-3mm}
\end{table*}

\begin{figure}[h!]
\centering
\includegraphics*[width=50mm]{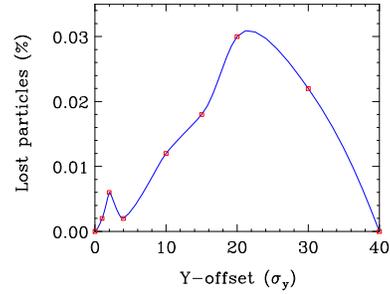}
\vspace{-3mm}
\caption{Particle loss in the extraction line vs. $\Delta y$ offset.}
\label{partloss-offy-H490}
\vspace{-3mm}
\end{figure}

As a result of the increased energy spread and vertical beam size, one can
expect higher beam loss in the extraction line for $\Delta y$ offsets
near 15 to 20$\sigma_y$.  The results of particle loss versus $y$-offset are
shown in Fig.~\ref{partloss-offy-H490}.  The maximum loss occurs at $\Delta
y\!=\!20\sigma_y$ and amounts to 0.7~kW of power loss with 15 lost
particles out of $5\!\times\!10^4$ in tracking.  Some irregularities in
Fig.~\ref{partloss-offy-H490} are due to low statistics of the lost
particles.  As expected, tracking with the horizontal IP offsets showed no
particle loss.

More detailed analysis of the beam loss with $\Delta y\!=\!20\sigma_y$
offset revealed that all of the particles except one were lost in the
vertical plane.  This indicates that the blow up of vertical emittance may
be the source of the particle loss.

Our expectation is that the power loss caused by IP offset, even at the
maximum value of 0.7~kW, can be safely disposed.  In practice, the losses
should be much lower since the beam offset will be controlled at the level
of $1\sigma_y$ for a maximum luminosity.  We expect that with such control
of the vertical offsets the power loss will be on the order of
$\sim$0.05~kW for the Stage 1 parameters.

\vspace{-1mm}
\section{ANGULAR OFFSET AT IP}

Disrupted beam distributions with angular offset at IP were generated in
the GUINEA--PIG code by changing the nominal initial angle at IP by
$\pm\frac{1}{2}\Delta x^\prime$ or $\pm\frac{1}{2}\Delta y^\prime$ in the
two beams.  The resultant distributions of $1\!\times\!10^5$
particles were then tracked in the extraction line.

\begin{figure}[t]
\centering
\includegraphics*[width=50mm]{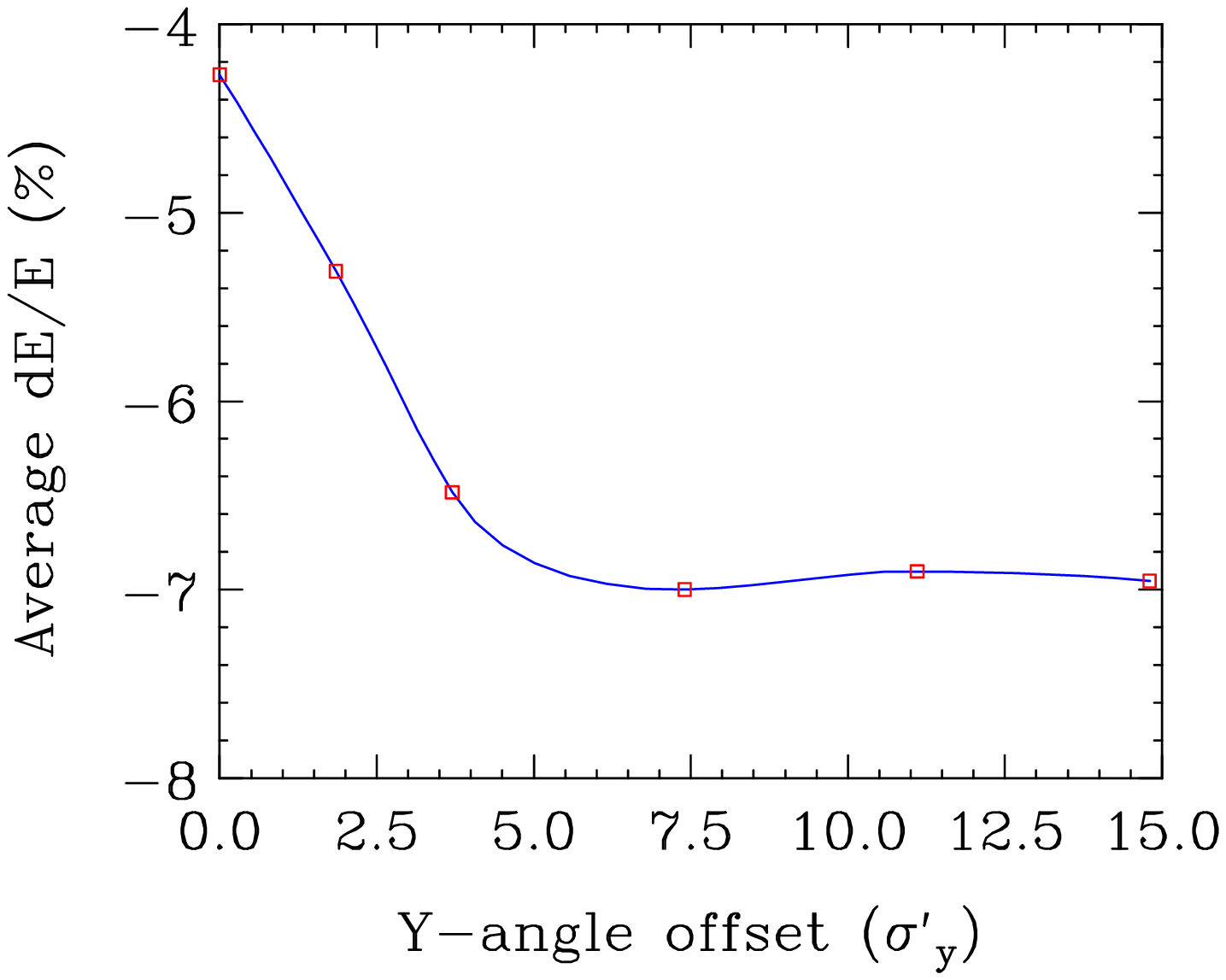}
\vspace{-3mm}
\caption{Average disrupted $\frac{\Delta E}{E}$ at IP vs. $\Delta y^\prime$.}
\label{deave-offangy-H490}
\vspace{-0mm}
\end{figure}

\begin{figure}[t]
\centering
\includegraphics*[width=50mm]{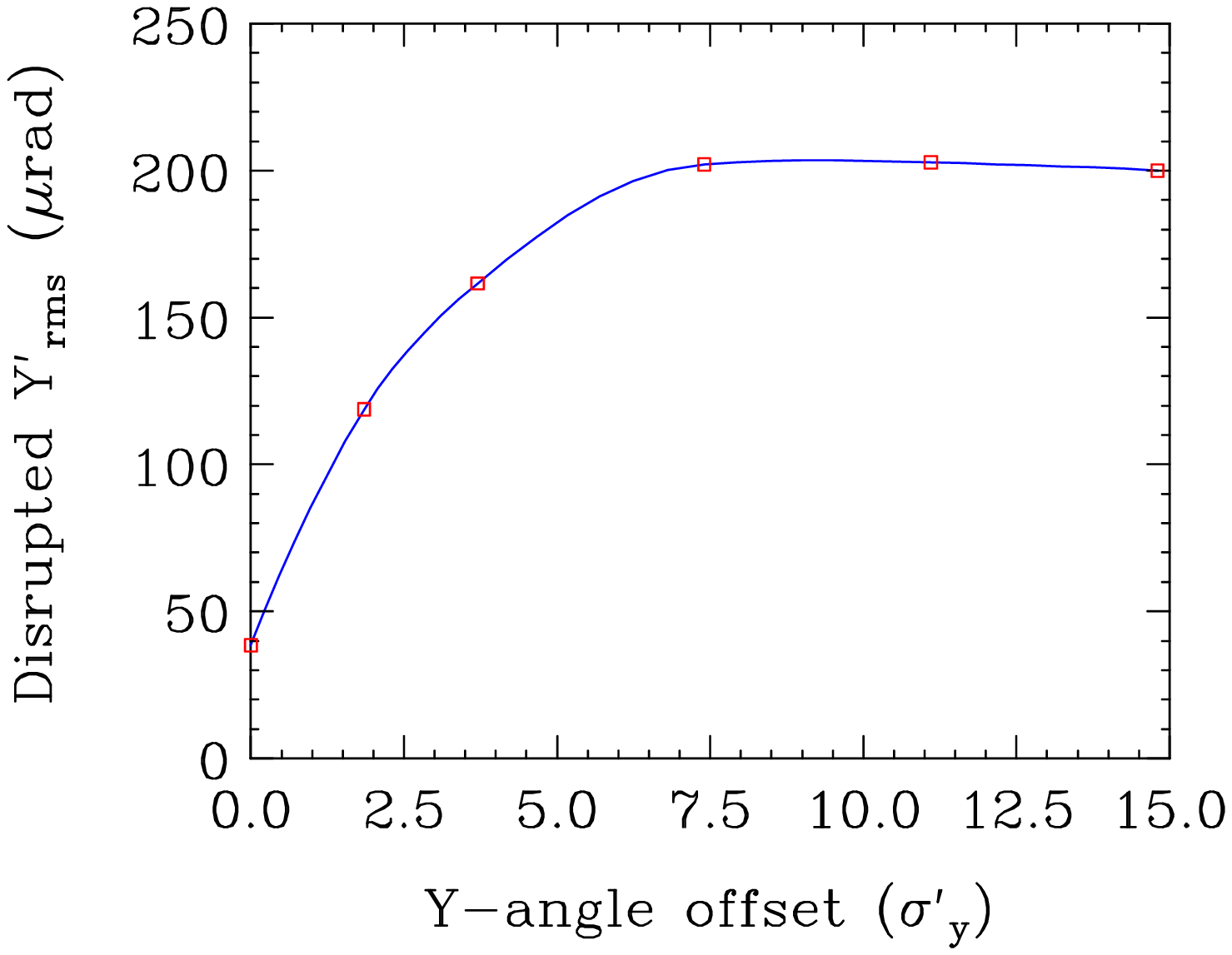}
\vspace{-3mm}
\caption{Disrupted $rms$ vertical divergence at IP vs. $\Delta y^\prime$.}
\label{sigpy-offangy-H490}
\vspace{-0mm}
\end{figure}

\begin{figure}[h!]
\centering
\includegraphics*[width=50mm]{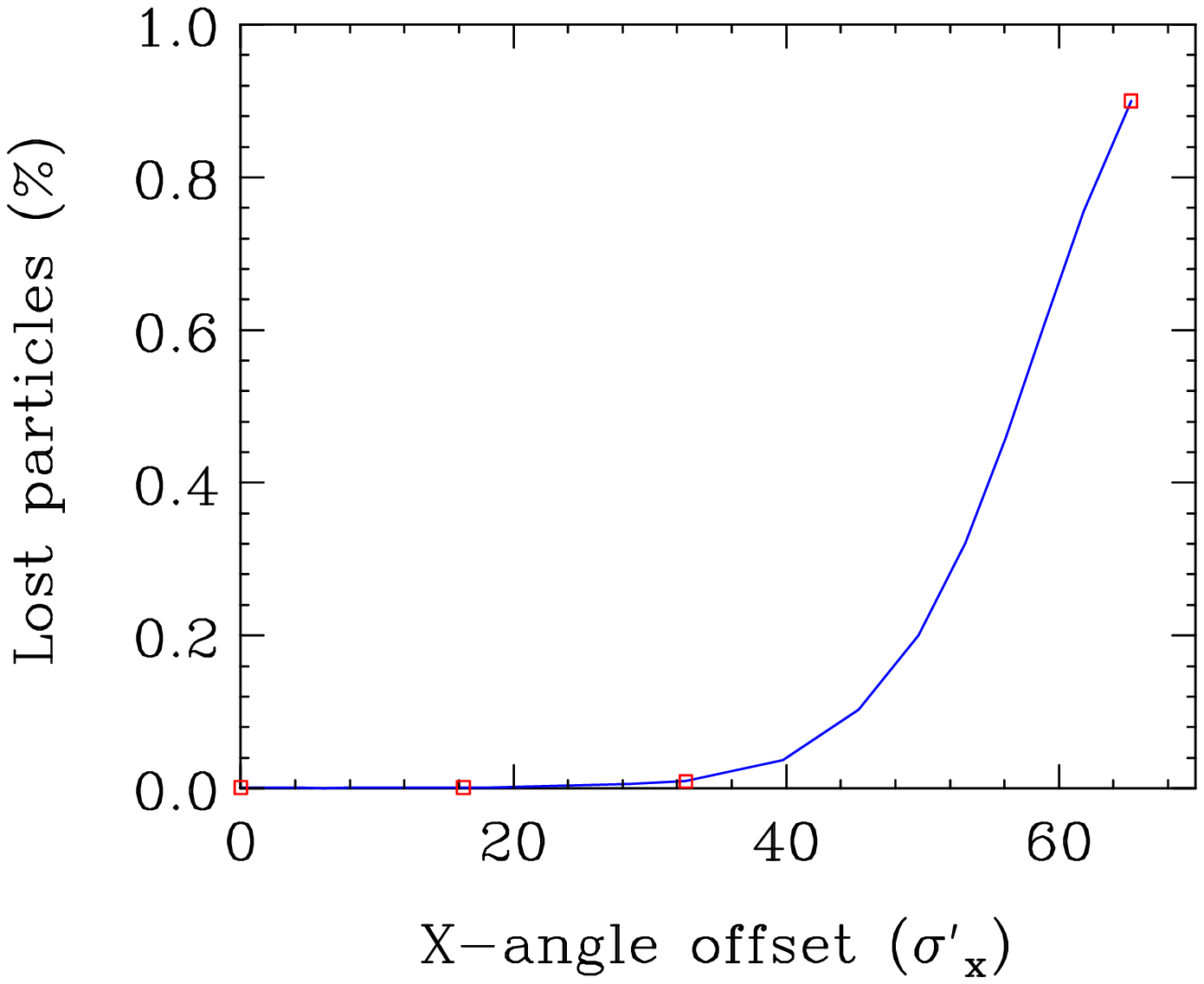}
\vspace{-3mm}
\caption{Particle loss in the extraction line vs. $\Delta x^\prime$ offset.}
\label{partloss-offangx-H490}
\vspace{-3mm}
\end{figure}

\begin{figure}[t]
\centering
\includegraphics*[width=50mm]{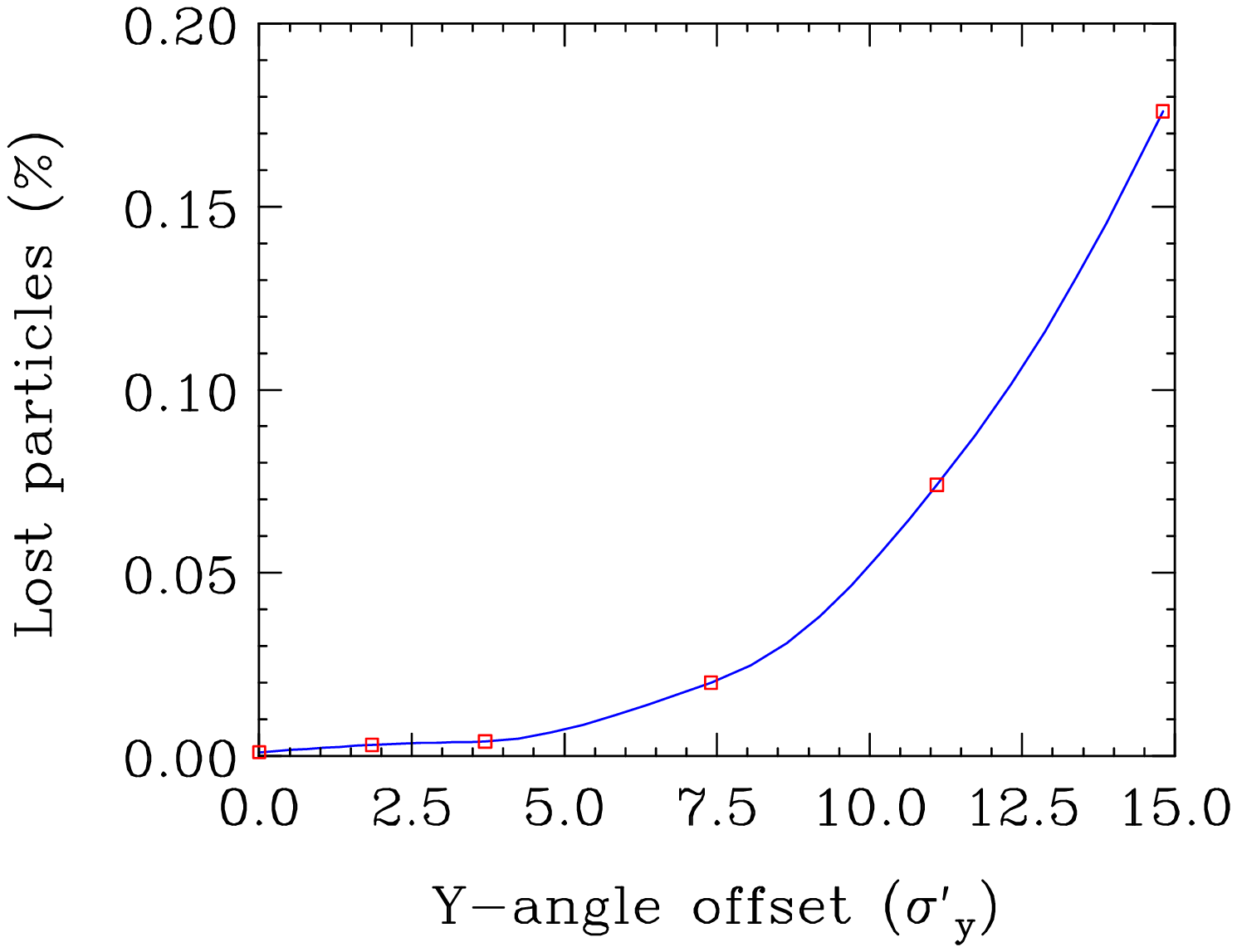}
\vspace{-3mm}
\caption{Particle loss in the extraction line vs. $\Delta y^\prime$ offset.}
\label{partloss-offangy-H490}
\vspace{-3mm}
\end{figure}

As in the case of IP position offset, a vertical angular offset results in
somewhat increased beam energy spread and vertical divergence at IP as
shown in Fig.~\ref{deave-offangy-H490} and~\ref{sigpy-offangy-H490}, while
the effect of a horizontal angle is small.  Note that angular offset in
Fig.~\ref{deave-offangy-H490} and~\ref{sigpy-offangy-H490} is normalized to
$\sigma^{\prime}_{x,y}$, the undisrupted initial divergence at IP.
Particle tracking in the extraction line showed, however, that large offset
angles have much stronger effect on the beam loss than the position offset.
The particle loss versus $\Delta x^\prime$ and $\Delta y^\prime$ is shown
in Fig.~\ref{partloss-offangx-H490} and~\ref{partloss-offangy-H490}.  The
large particle losses are caused by the increased beam orbit oscillations
in the extraction line proportional to the IP angular offsets.

For a beam power loss below 1~kW at a center-of-mass energy of 500~GeV, the
particle loss needs to be lower than 0.05\%.  According to
Fig.~\ref{partloss-offangx-H490} and~\ref{partloss-offangy-H490}, the
corresponding maximum full angular offsets are about 40$\sigma^{\prime}_x$
(1.2~mrad) and 10$\sigma^{\prime}_y$ (0.25~mrad) for $x$ and $y$
planes, respectively.  Note that the beam loss is strongly reduced at
smaller angles.  However, to avoid unnecessary large beam loss, the angular
offsets should be kept well below the above tolerances.

\vspace{-1mm}
\section{CONCLUSION}

The beam position and angular offsets at IP may increase the beam
disruption.  For the Stage 1 (500 GeV {\it cms}) NLC parameters, the beam
vertical offset may result in up to 0.7~kW beam loss in the extraction
line.  In practice, the losses will be much lower for reasonably controlled
IP beam positions.  The angular offsets should be kept lower than
40$\sigma^{\prime}_x$ (1.2~mrad) and 10$\sigma^{\prime}_y$ (0.25~mrad) in
$x$ and $y$ planes for a power loss below 1~kW.

\end{document}